\documentclass{ametsocV6.1}

\usepackage{hyperref}
\hypersetup{
	breaklinks = true,
	hidelinks  = true
}

\title{Dependence of convective precipitation extremes on near-surface relative humidity}

\authors{Robert J. van der Drift\aff{a}\correspondingauthor{Robert van der Drift, rvddrift@mit.edu}
Paul A. O'Gorman\aff{a}}

\affiliation{\aff{a}{Department of Earth, Atmospheric and Planetary Sciences, Massachusetts Institute of Technology}}

\abstract{Precipitation extremes produced by convection have been found to intensify with near-surface temperatures at a Clausius-Clapeyron rate of $6$ to $7\%$ K$^{-1}$ in simulations of radiative-convective equilibrium (RCE). However, these idealized simulations are typically performed over an ocean surface with a high near-surface relative humidity (RH) that stays roughly constant with warming. Over land, near-surface RH is lower than over ocean and is projected to decrease by global climate models. Here, we investigate the dependence of precipitation extremes on near-surface RH in convection-resolving simulations of RCE. We reduce near-surface RH by increasing surface evaporative resistance while holding free-tropospheric temperatures fixed by increasing surface temperature. This ``top-down'' approach produces an RCE state with a deeper, drier boundary layer, which weakens convective precipitation extremes in three distinct ways. First, the lifted condensation level is higher, leading to a small thermodynamic weakening of precipitation extremes. Second, the higher lifted condensation level also reduces positive buoyancy in the lower troposphere, leading to a dynamic weakening of precipitation extremes. Third, precipitation re-evaporates more readily when falling through a deeper, drier boundary layer, leading to a substantial decrease in precipitation efficiency. These three effects all follow from changes in near-surface relative humidity and are physically distinct from the mechanism that underpins the Clausius-Clapeyron scaling rate. Overall, our results suggest that changes in relative humidity must be taken into account when seeking to understand and predict changes in convective precipitation extremes over land.}

\begin{document}

%% Necessary!

%% Do not count words:
\maketitle

%%%%%%%%%%%%%%%%%%%%%%%%%%%%%%%%%%%%%%%%%%%%%%%%%%%%%%%%%%%%%%%%%%%%%
% SIGNIFICANCE STATEMENT/CAPSULE SUMMARY
%%%%%%%%%%%%%%%%%%%%%%%%%%%%%%%%%%%%%%%%%%%%%%%%%%%%%%%%%%%%%%%%%%%%%
%
% If you are including an optional significance statement for a journal article or a required capsule summary for BAMS 
% (see www.ametsoc.org/ams/index.cfm/publications/authors/journal-and-bams-authors/formatting-and-manuscript-components for details), 
% please apply the necessary command as shown below:

%
% Significance Statement (all journals except BAMS)
%

%TC:ignore
\statement
Thunderstorms and other types of convection produce very heavy rainfall in many regions on Earth. In this paper, we ran a computer model to show that when relative humidity near the surface is reduced, convection produces weaker rainfall rates. This happens for three reasons: the updrafts in the storms are weaker, there is less cloud-water because the cloud base is higher, and less of that cloud-water makes it to the surface as rainfall. This is an important finding because we expect that relative humidity will decrease over many land regions as the climate warms, possibly seasonally offsetting some of the impact on rainfall extremes of increases in the absolute amount of water vapor.

\section{Introduction}

The heaviest of precipitation events are affected by climate change through a ``thermodynamic contribution'' related to increasing water vapor, a ``dynamic'' contribution related to changes in vertical velocities, and by changes in ``precipitation efficiency,'' the fraction of condensed water vapor that actually reaches the surface as precipitation \citep{o2015precipitation}. For convective precipitation extremes, the thermodynamic contribution approximately follows a Clausius-Clapeyron scaling rate of $6$ to $7\%$ per K of surface warming \citep{muller2011intensification,romps2011response,abbott2020convective}. Precipitation extremes scale near the Clausius-Clapeyron rate in global climate model (GCM) projections at the global scale, albeit with large uncertainties in the tropics  \citep[e.g.,][]{ogormanschneider,kharin2013changes}, in some regional studies with convection-resolving models (CRMs) \citep[e.g.,][]{ban2015heavy,prein2017future}, and in globally-aggregated observations over land \citep{westra13}. Evidence exists, however, from observations \citep{fowler2021anthropogenic}, GCM projections \citep{pfahl2017regional,williams2022summer}, and regional CRM studies \citep{lenderink2021superCC} that precipitation extremes may regionally and seasonally respond to climate change at a rate that deviates from Clausius-Clapeyron scaling.

\citet{williams2022summer}, in particular, found a seasonal contrast in the scaling rates of precipitation extremes with climate warming across simulations from the Coupled Model Intercomparison Project, Phase 5 (CMIP5) \citep{CMIP5}. They found that over midlatitude land in the Northern Hemisphere, dynamic contributions to precipitation extremes were near-zero in the winter but negative in the summer. This negative dynamic contribution is likely related to convection, since convective precipitation extremes are common in the summer. \citet{williams2022summer} further found a correlation between these summertime negative dynamical contributions and a decrease in summertime near-surface relative humidity (RH), suggesting that convective precipitation extremes respond dynamically to decreases in near-surface RH. 

Near-surface RH is often thought to influence precipitation extremes and their response to climate change through the Clausius-Clapeyron rate. Only when RH stays constant will near-surface specific humidity scale one-to-one with saturation specific humidity and follow the Clausius-Clapeyron rate. With this in mind, a number of papers have opted to scale precipitation extremes against dew-point temperature instead of temperature, arguing that a direct measure of atmospheric moisture content should produce a scaling, absent other effects, that follows the Clausius-Clapeyron rate \citep{lenderink2010dewpoint,lenderink2011scaling,lepore2015temperature,barbero2018dewpoint,lenderink2021superCC}. However, arguments in favor of such a ``dew-point scaling'' approach over a more traditional ``temperature scaling'' implicitly assume that RH \textit{only} matters for its influence on the thermodynamic contribution to changes in precipitation extremes. By finding a relationship between RH and a dynamic contribution, \citet{williams2022summer} have called this assumption into question.

Near-surface RH is expected to decrease over land in response to anthropogenic climate change for several reasons. First, RH is expected to decrease over land because water vapor over land is influenced by moisture transport from over ocean, while at the same time ocean warming is weaker than the land warming. Thus, the source of water vapor from over ocean can't keep pace with the increasing saturation vapor pressure over land \citep{simmons2010,byrne2016understanding,byrne2018obs}. In addition, surface evapotranspiration rates provide a direct control on near-surface RH. Surface evapotranspiration and thus near-surface RH is reduced by a ``physiological forcing'' in which plant stomata close in response to higher atmospheric CO$_2$ levels \citep{cao2010physiological}, and this stomatal closure has been found to decrease mean precipitation in summer over the northern midlatitudes \citep{skinner17}. Lastly, decreases in soil moisture are also expected to influence near-surface RH \citep{berg2016sm,zhou2023intersm}. Comparison between observed and simulated trends in RH in recent decades shows that GCMs underestimate decreases in near-surface RH in arid and semi-arid regions \citep{simpson24}, which means that they may also underestimate any resulting impacts on precipitation in these regions.

In this paper, we investigate the sensitivity of convective precipitation extremes to near-surface RH in the simplest possible setting: a CRM run to a state of radiative-convective equilibrium (RCE). In regional simulations, across a wide range of relative humidities, CRMs have been demonstrated to reproduce observed precipitation extremes more reliably than models that use convective parameterizations \citep{lenderink2024}. In RCE simulations, CRMs are additionally useful because they allow for careful, controlled study of the physics underlying precipitation statistics and their response to different climate forcings. A number of studies of idealized CRM simulations of RCE have found that convective precipitation extremes scale quite close to the Clausius-Clapeyron rate in response to warming \citep{muller2011intensification,romps2011response}. Dynamic contributions to precipitation extremes remain relatively small in these idealized studies, even when convection is organized into squall lines by wind shear \citep{muller2013squall} or overturning structures in a channel domain \citep{abbott2020convective}. These particular studies also did not find large changes in precipitation efficiency, but \citet{singh2014microphysics} did find that precipitation efficiency decreased in colder RCE states due to microphysical effects. Several other idealized CRM studies have diagnosed the importance of various physical processes in setting the precipitation efficiency for both mean and extreme precipitation \citep{lutsko2018increase,da2021significant,abramian2023squall,langhans2015lagrangian}. However, the RCE studies cited above have all used an ocean surface as a bottom boundary condition, and so the influence of near-surface RH on precipitation extremes in states of RCE has remained relatively unexplored. 

We are aware of two CRM studies of the effect of overall surface dryness on convective intensity in RCE: \citet{hansen2015bowen} and \citet{sarbengthesis}. These studies were motivated by observational evidence that convection is more intense over land than over ocean \citep{zipser2006intensity}. Both studies found that the maximum updraft velocity does not increase with a higher Bowen ratio (less evaporative surface), which suggests that the land-ocean contrast in convective intensity is not due to the contrast in surface dryness. \citet{sarbengthesis} even found weakening in updrafts in the lower free troposphere as the surface dries, which could be consistent with the negative dynamic contribution to precipitation extremes found by \citet{williams2022summer} in response to lower near-surface RH, especially given that condensation rates are sensitive to updraft velocities in the lower troposphere where saturation vapor pressures are relatively high.

To modify near-surface RH in our simulations, we introduce a ``vegetative'' evaporative resistance parameter similar to \citet{betts2000idealized} and inspired by the effects of stomatal closure on surface relative humidity. As the climate changes, decreases in near-surface RH occur alongside increases in near-surface temperatures, such that the near-surface moist static energy and free-tropospheric temperatures (which are convectively coupled at equilibrium and under the influence of larger-scale dynamics) tend not to change as much \citep[e.g.,][]{byrne2013gcm,berg2016sm}. Thus we follow the general approach of \citet{hansen2015bowen} and \citet{sarbengthesis} by holding the free-tropospheric temperature fixed as the surface dries, specifically using the relaxation procedure introduced by \citet{sarbengthesis}. For simplicity, we do not parameterize the effects of changes in large-scale dynamics or include the diurnal cycle, both of which should be considered in future work.

Section 2 describes the vegetative resistance parameter, the relaxation procedure, and the model simulations more generally. Section 3 presents an overview of the mean RCE state achieved by varying the vegetative resistance and the fundamental result of this paper: that precipitation extremes vary substantially with near-surface RH, and that the mechanisms involves changes in dynamics and precipitation efficiency rather than the thermodynamic contribution that gives rise to the Clausius-Clapeyron scaling rate of precipitation extremes. Sections 4 and 5 explain this dependence in more detail. Specifically, Section 4 shows that a higher lifted condensation level weakens convective updrafts in the lower troposphere, while Section 5 diagnoses changes in precipitation efficiency in terms of cloud microphysics and re-evaporation. Section 6 provides a concluding discussion, highlighting the implications of a large sensitivity of precipitation extremes to near-surface RH.

\section{Model and simulations}

\subsection{Convection-resolving model and basic setup}

We use the System for Atmospheric Modeling (SAM), version 6.11 \citep{SAM}. All simulations were run with a $1$ km horizontal grid spacing in a $128\times 128$ km$^2$ domain, and with $64$ vertical levels. Vertical spacing starts at $37.5$ m near the surface and increased steadily until the model top at a height of $27$ km. Above $16$ km, atmospheric motions are damped in a sponge layer. SAM was run with its own one-moment microphysics parameterization. No diurnal cycle was simulated; instead, a constant zenith angle of $42.3^\circ$ was used. This zenith angle, along with a solar constant set to $565$ W/m$^2$, produces Earth's equatorial annual average of insolation weighted by the cosine zenith angle, following the recommendations of \citet{cronin2014zenith}. 

\subsection{Vegetative evaporative resistance and surface temperature}

We modify near-surface RH by altering the rate of surface evaporation. To accomplish this, a free parameter, the ``vegetative resistance'' $r_v$, was introduced in SAM's equation for the surface latent heat flux:
\begin{equation}
    \mathrm{LHF} = \rho L_v\frac{\Delta q}{r_{ae} + r_v},
\end{equation}
where $\rho$ is near-surface air density, $L_v$ is the latent heat of vaporization, $\Delta q$ is the difference between near-surface mixing ratio $q_v$ and the saturation mixing ratio at surface (skin) temperature $T_s$, and $r_{ae} = (C_e U)^{-1}$ is an ``aerodynamic resistance.'' For the aerodynamic resistance, $C_e$ is a unitless exchange coefficient determined by Monin-Obukhov similarity theory and $U$ is the near-surface windspeed. When calculating surface heat fluxes, SAM sets $U$ to have a minimum value of $1$ m s$^{-1}$ to account for unresolved gusts.

We model the surface to be horizontally homogeneous, so that there are no spatial variations in $r_v$ or in the surface temperature $T_s$. The surface is an ocean when $r_v = 0$ s m$^{-1}$ and surface evaporation rates decrease as $r_v$ is increased (all else held equal). Modifications to evaporation through $r_v$ affect the surface energy budget, and so $T_s$ may not stay constant as $r_v$ is varied. An intuitive approach to determining a value for $T_s$, given a value of $r_v$, is to simply to solve the surface energy budget until equilibrium is achieved, as has been done by some past RCE studies (e.g., \citet{romps2011response}). Simulations using this approach (not shown) reached a state of equilibrium with \textit{lower} $T_s$ at large $r_v$, causing the free troposphere to cool substantially.\footnote{As $r_v$ increased, the free troposphere in these simulations dried and $T_s$ cooled in order to maintain the same outgoing longwave radiation. This kind of radiative response to $r_v$ (and similar parameters) has been found previously in GCM studies that varied evaporation rates via fractional coverage of land continents vs. ocean \citep{lague2021config,lague2023config}.}
Such a free-tropospheric cooling is inconsistent with the constraint of weak temperature gradients (WTG) in the tropics. Instead, we take a ``top-down'' perspective on the controls on land temperatures at climate equilibrium, which argues that free-tropospheric temperatures over land are strongly coupled vertically to surface temperature and moisture in convecting regions by moist adiabatic lapse rates, and also strongly coupled horizontally to free-tropospheric temperatures over ocean by horizontal advection and gravity wave dynamics. This perspective has previously been used to explain the land-ocean warming and moistening contrasts under climate change
\citep[e.g.,][]{joshi2008contrast,byrneogormancontrast}, and it implies that free-tropospheric temperatures over land should not necessarily change in response to a change in surface evapotranspiration.
%This top-down model has been found to be empirically accurate in both the tropics and midlatitudes \citep{byrne2013gcm,byrne2018obs}, even though the assumption of constant free-tropospheric temperatures was based on the tropical weak temperature gradient (

With this perspective in mind, we use a method devised by \citet{sarbengthesis}, which adjusts $T_s$ so that horizontal-mean temperature $T$ is nudged towards a reference profile $T_{\text{ref}}$ within a specified pressure layer between $p_\mathrm{lower}$ and $p_\mathrm{upper}$. That is, we evolve $T_s$ forward in time using
\begin{equation}
    \frac{dT_s}{dt} = \frac{1}{\tau \Delta p}\int_{p_{\mathrm{upper}}}^{p_{\mathrm{lower}}} (T_{\text{ref}}(p) - T(p,t))\ dp,\label{eq:fft_relax}
\end{equation}
where $\Delta p = p_{\mathrm{lower}} - p_{\mathrm{upper}}$ is the thickness of the layer and $\tau$ is a relaxation timescale. This implementation differs from \citet{sarbengthesis} in two ways. First, the integral was evaluated in pressure coordinates, not height coordinates, with $p_{\mathrm{lower}} = 600$ hPa and $p_{\mathrm{upper}} = 400$ hPa. Second, a longer relaxation timescale of $\tau = 3.6$ days was used instead of $\tau = 6$ hr. Both of these modifications dampened oscillations in $T_s$ that appeared in initial attempts to apply this adjustment. The reference profile was calculated from a simulation with $r_v=0$ as described in the next subsection.

An alternative approach would be to parameterize WTG dynamics by introducing a large-scale vertical velocity that prevents large changes in free-tropospheric temperature  \citep{sobel2000modeling,raymond2005modeling}. The role of changes in large-scale vertical velocities is an important topic for future work, but here we focus on the simplest case of RCE.

\subsection{Simulations}

In total, $5$ simulations were run. Each simulation is associated with a different vegetative resistance $r_v$: $0$, $200$, $500$, $1000$, and $2000$ s m$^{-1}$. The same reference profile $T_{\text{ref}}(p)$ was used to determine a surface temperature $T_s(r_v)$ for each simulation. The reference profile was calculated by first running SAM in an ``ocean RCE'' configuration: $r_v = 0$ s m$^{-1}$ and $T_s = T_{s,o} = 300$ K. This ocean RCE simulation was run for 50 days, and $T_{\text{ref}}$ was calculated by averaging horizontally and over the last 10 days of the simulation.

Once $T_{\text{ref}}$ was determined, the following procedure was used for every simulation (including $r_v = 0$ s m$^{-1}$). First, SAM was run for $40$ days, with $T_s$ evolved forward in time following Equation (\ref{eq:fft_relax}). If, averaged horizontally and over the last $10$ days of this simulation, the vertical average between $p_{\mathrm{low}}$ and $p_{\mathrm{upp}}$ of temperature and the reference profile $T_{\text{ref}}$ were within $0.1$ K, then the average value of $T_s$ over those $10$ days was saved. Otherwise, the simulation was run for $10$ more days and the procedure was repeated. Once a value of $T_s$ was saved, SAM was re-run from rest with this fixed value of $T_s$ for $60$ days. The last $30$ days of those fixed-$T_s$ simulations were used for all of the analysis presented below; during this $30$ day period, instantaneous snapshots from SAM were saved every $3$ hours. Table \ref{tab:Ts} reports the equilibrium values of $T_s$, as well as the horizontal- and time-average near-surface temperature, near-surface RH, near-surface $q_v$, and surface precipitation rates for all simulations. The values of $r_v$ were spaced further apart at large values of $r_v$, so that for each increment in $r_v$,  $T_s$ increased by approximately $2$ K and RH decreased by approximately $7-8\%$ (Table \ref{tab:Ts}). 
%Unless otherwise stated, the results in this paper focus on the $T_{s,o} = 300$ K simulations, however *Supplemental figure shows that our main finding holds when $T_{s,o}$ is varied (*SF: versions of Fig. \ref{fig:frac_tree} for each $T_{s,o}$).

\begin{table}
	\begin{center}
		\begin{tabular}{c|c|c|c|c|c}
            %\multicolumn{6}{c@{}}{$\mathbf{T_{s,o} = 300}$\textbf{ K}}\\
            $r_v$ (s m${-1}$) & $T_s$ (K) & near-sfc $T$ (K) & near-sfc RH (\%) & near-sfc $q_v$ (g kg${-1}$)& $P$ (mm day$^{-1}$) \\ \hline
			$0$    & $300.0$ & $296.9$ & $75.2$ & $14.0$ & $2.8$\\ 
			$200$  & $301.9$ & $297.7$ & $68.5$ & $13.4$ & $2.4$\\ 
			$500$  & $303.9$ & $298.7$ & $61.3$ & $12.8$ & $2.1$\\ 
			$1000$ & $306.0$ & $299.9$ & $53.1$ & $11.9$ & $1.6$\\ 
			$2000$ & $308.3$ & $301.4$ & $44.3$ & $10.9$ & $1.2$\\
		\end{tabular}

		\caption{Horizontal- and time-mean variables for each $r_v$ simulation: surface temperature $T_s$ (determined by the relaxation procedure); near-surface air temperature $T$, RH, and mixing ratio $q_v$; and surface precipitation rate $P$. Note that ``near-surface'' refers to the lowest model level (40 m).}\label{tab:Ts}
	\end{center}
\end{table}

\section{Response to changes in surface dryness}\label{sec:response}

\subsection{Response of mean climate}

\begin{figure}[ht]
	\centering\includegraphics[scale=0.7]{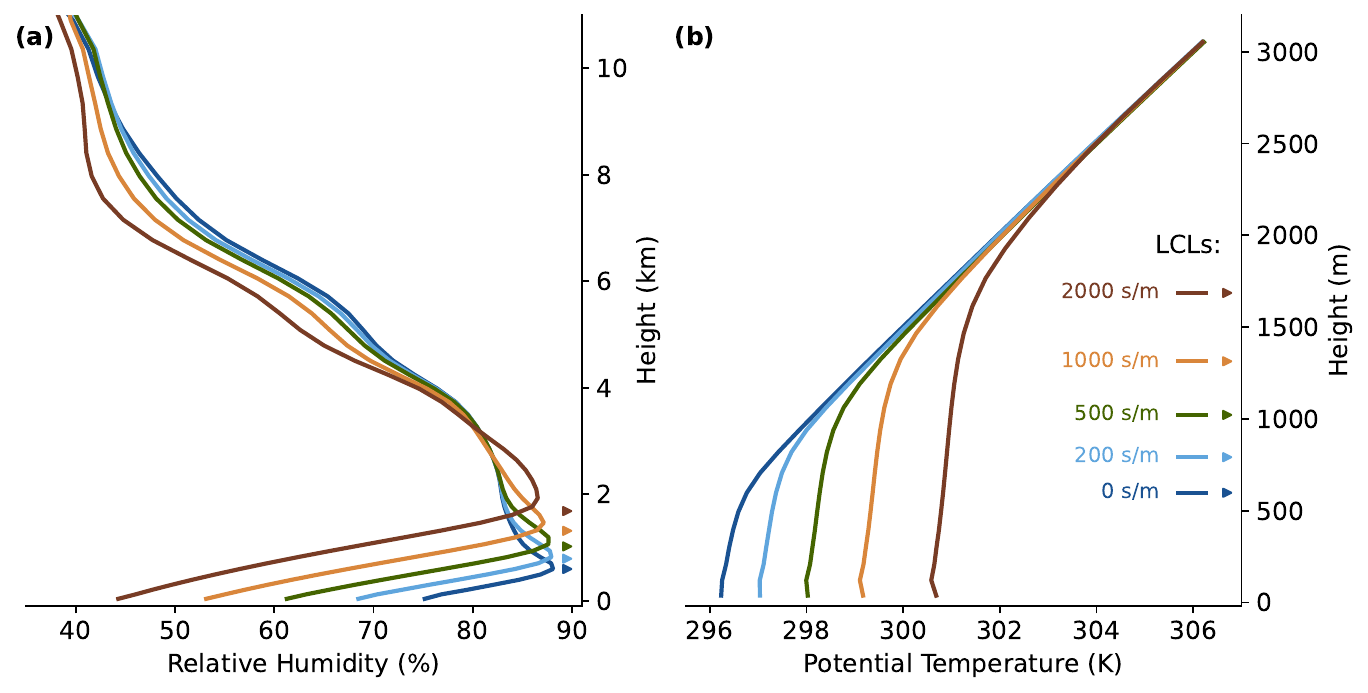}
	\caption{Horizontal- and time-mean: (a) near-surface relative humidity and (b) potential temperature versus height for simulations with varied vegetative resistance $r_v$ as labeled in the legend. The legend is spaced vertically so that each horizontal line is at the LCL (denoted on the $y$-axis of both panels with triangles). The LCL is calculated following \citet{romps2017LCL} and using near-surface horizontal- and time-mean relative humidity and temperature. Note the different vertical axis ranges in panels (a) and (b).}\label{fig:th_rh}
\end{figure}

As $r_v$ increases, the boundary layer changes in three ways. First, near-surface RH and specific humidity decrease with increasing $r_v$ (Table \ref{tab:Ts} and Fig. \ref{fig:th_rh}a). Second, temperatures increase in the boundary layer even as temperature stays constant in the free troposphere (Table \ref{tab:Ts} and Fig. \ref{fig:th_rh}b). Third, the lifted condensation level (LCL), calculated analytically \citep{romps2017LCL} using horizontal- and time-mean near-surface air properties, rises as $r_v$ increases (Fig. \ref{fig:th_rh}b). If we consider the $r_v=0$ s m$^{-1}$ simulation as ``ocean'' and the $r_v>0$ s m$^{-1}$ simulations as ``land,'' then these tendencies are consistent with the top-down perspective on land-ocean contrasts of \citet{byrneogormancontrast} (c.f. their Fig. 1), since the surface temperature over land must be higher given the same free-tropospheric temperatures over land and ocean, moist- adiabatic lapse rates above the LCL and dry adiabatic lapse rates below the LCL, and a higher LCL over land. One difference is that lapse rates below the LCL in our simulations shown in Fig. \ref{fig:th_rh}b are not quite dry adiabatic because of precipitation-driven cold pools.

The reference temperature profile, $T_{\text{ref}}$, calculated over an ocean surface ($r_v = 0$ s/m), is closely followed through the free troposphere in simulations with $r_v > 0$ s/m. Thus, in conjunction with the relaxation procedure, $r_v$ acts as a control on the near-surface RH without modifying free-tropospheric temperatures. Relative humidity also decreases somewhat in the free troposphere with increasing $r_v$.

\subsection{Response of precipitation extremes}

Every horizontal gridpoint and time in a given simulation has its own value of the instantaneous precipitation rate $P$ (saved directly) and the vertically-integrated condensation rate $C$ (calculated as described below from Equation (\ref{eq:cond_int})). We characterize extreme precipitation $P_e$ by the average of $P$ over all values at and above the $99.9$th percentile of $P$. Similarly, extreme condensation $C_e$ is the average of $C$ over all values at and above the $99.9$th percentile of $C$. Precipitation efficiency $\epsilon_p$ is defined as the ratio $P_e/C_e$, and thus involves different sets of gridpoints and times for $P_e$ and $C_e$, following the approach of \citet{singh2014microphysics}, \citet{abbott2020convective}, and \citet{da2021significant} which recognizes that the different variables peak at different points in the convective lifecycle. 

\begin{figure}[ht]
    \centering\includegraphics[width=\linewidth]{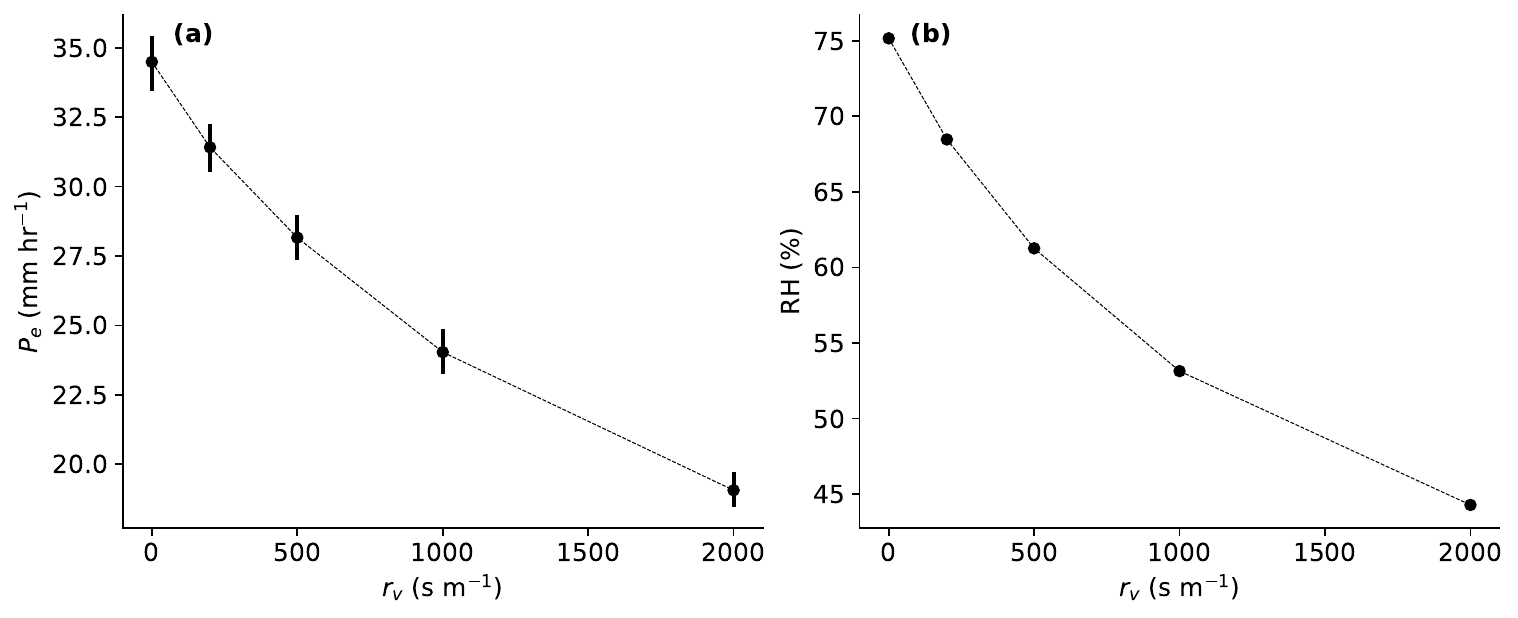}
    \caption{(a) $99.9$th-percentile threshold average of instantaneous precipitation rates across the simulations. Error bars show the $5$th to $95$th percentile of the threshold average calculated using block bootstrapping. (b) Horizontal- and time-mean relative humidity across the simulations.}
    \label{fig:pe_rh}
\end{figure}

We find that precipitation extremes weaken substantially as the surface dries and RH decreases (Fig. \ref{fig:pe_rh}). When $r_v$ increases from $0$ s m$^{-1}$ to $2000$ s m$^{-1}$, precipitation extremes decrease from $34.5$ mm hr$^{-1}$ to $19.2$ mm hr$^{-1}$ (Fig. \ref{fig:pe_rh}a), a $57\%$ fractional decrease. Over the same range of $r_v$, near-surface RH decreases from $75.2\%$ to $44.3\%$ (Fig. \ref{fig:pe_rh}b). This is an absolute decrease in RH of $31$ percentage points ($\%$pt), or a fractional decrease of $52\%$. Thus, we calculate (following the methodology described in Section \ref{sec:response}\ref{sec:decomp}) the $P_e$ scaling rate between our wettest and driest simulations either as a $1.9\%$ per $\%$pt (absolute) increase in RH or as a $1.1\%$ per $\%$ (fractional) increase in RH. This scaling rate is the same sign as, but half the magnitude of, the roughly $2\%$ per $\%$ scaling rate found by \citet{williams2022summer} for the dynamical contribution to changes in precipitation extremes over northern hemisphere midlatitude land in the summertime (c.f. their Fig. 3). 

Note that this scaling rate is computed between simulations with a large difference in RH. The scaling rate against absolute changes in RH is smaller over wetter surfaces ($1.4\%$ per $\%$pt between $r_v = 0$ s m$^{-1}$ and $200$ s m$^{-1}$) and larger over drier surfaces ($2.5\%$ per $\%$pt between $r_v = 1000$ s m$^{-1}$ and $2000$ s m$^{-1}$). In contrast, the scaling rate against fractional changes in RH varies less between wetter surfaces ($1.0\%$ per $\%$ between $r_v = 0$ s m$^{-1}$ and $200$ s m$^{-1}$) vs. drier surfaces ($1.2\%$ per $\%$ between $r_v = 1000$ s m$^{-1}$ and $2000$ s m$^{-1}$). Given this reduced variability, we present our results in terms of scaling rates against fractional changes in RH across our full range of simulations (unless otherwise stated). This is also consistent with \citet{williams2022summer}, who report fractional changes in RH (A. Williams, 2024, personal communication).

Although we focus on precipitation extremes in this paper, we have also found that the fractional decrease in $P_e$ is nearly identical to a $58\%$ decrease, from $2.8$ mm day$^{-1}$ to $1.2$ mm day$^{-1}$, in the horizontal- and time-mean precipitation rate (Table \ref{tab:Ts}). This is different from CRM and GCM simulations that either warm the surface or increase CO$_2$ concentrations, which find that mean precipitation rates are energetically constrained to scale with warming below the CC scaling rate that precipitation extremes approximately follow \citep{allen2002constraints,heldsoden,ogormanschneider,muller2011intensification}.

\subsection{Decomposition of response of precipitation extremes}\label{sec:decomp}

High percentile instantaneous surface precipitation rates are associated with deep convection wherein strong updrafts condense moisture. SAM does not explicitly calculate condensation rates, and we want to decompose changes in the condensation rate into contributions from different physical factors. Therefore, we calculate the column-integrated condensation rate $C$ as 
\begin{equation}
    C = \int_{z_{\mathrm{LCL}}}^{z_t}-\left(\frac{dq_v^*}{dz}\right)_{\mathrm{ma}} \rho\tilde{w}\ dz,\label{eq:cond_int}
\end{equation}
where $z_{\mathrm{LCL}}$ is the height of the LCL calculated following \citet{romps2017LCL} using near-surface values of the column, $z_t = 14$ km is a fixed upper height, $w$ is vertical velocity, $\tilde{w}\equiv\max(0,w)$ is the updraft speed (i.e., excluding downdrafts), $q_v^*$ is the saturation mixing ratio (a function of only temperature $T$ and pressure $p$), and the subscript $\mathrm{ma}$ indicates that the derivative $dq_v^*/dz$ is calculated following a local moist adiabatic lapse rate. This definition of the condensation integral differs from the common definition \citep[e.g.,][]{ogormanschneider}. First, a lower bound of $z = z_{\mathrm{LCL}}$ is used instead of $z = 0$. Second, $\tilde{w}$ is used instead of $w$. Both of these choices are made on a physical basis: only upward velocities drive condensation, and convective clouds do not typically extend to the surface. The alternative choice of using $w$ instead of $\tilde{w}$ gives an approximate expression for net condensation (i.e., condensation from updrafts minus re-evaporation from downdrafts), but we include only updrafts so that the effect of re-evaporation is fully included in the precipitation efficiency. Also, given that the LCL rises with increasing $r_v$ (Fig. \ref{fig:th_rh}), it is important to diagnose the effect that the LCL has on condensation rates in order to correctly associate thermodynamic contributions, dynamic contributions, and changes in precipitation efficiency with the appropriate underlying mechanisms.

Letting $\delta(\cdot)$ denote the difference in a variable between two climate states and $\overline{(\cdot)}$ denote the average value of that variable between two climate states, then the relation $P_e=\epsilon_p C_e$ allows for fractional changes in precipitation to be decomposed in terms of fractional changes in efficiency and condensation:
\begin{equation}
    \frac{\delta P_e}{\overline{P_e}} = \frac{\delta\epsilon_p}{\overline{\epsilon_p}} + \frac{\delta C_e}{\overline{C_e}},\label{eq:frac_P}
\end{equation}
where we have neglected a nonlinear term (which is small for sufficiently close climate states). To minimize this nonlinear term, we calculate fractional changes of an extreme variable between adjacent simulations first (e.g., between $r_v = 0$ s m$^{-1}$ and $r_v = 200$ s m$^{-1}$), and then sum these together to get the total fractional change between the wettest and driest simulations.\footnote{This is, to good approximation, equal to the change in the logarithm of the extreme variable (e.g., $\delta\ln P_e$), which ensures that decompositions such as Equation (\ref{eq:frac_P}) are exact (e.g., $\delta\ln P_e = \delta\ln C_e + \delta\ln\epsilon_p$ without approximation). The advantage of our approach is that the decomposition into thermodynamic and dynamic contributions, Equation (\ref{eq:thermo_dynam}), cannot be written in terms of the logarithm of an extreme variable.} To get scaling rates, total fractional changes are normalized by the difference in the logarithm of horizontal- and time-mean $\mathrm{RH}$ between the wettest and driest simulations. 

Fractional changes in condensation may, in turn, be decomposed into thermodynamic and dynamic contributions by using Equation (\ref{eq:cond_int}). In this paper, thermodynamic contributions to condensation refer to changes in $(dq_v^*/dz)_\mathrm{ma}$, which is a function of $T$ and $p$, and also to changes in $z_{\text{LCL}}$, which is a function of near-surface $T$ and $\mathrm{RH}$ \citep{romps2017LCL}. Dynamic contributions to condensation refer to changes in $\rho\tilde{w}$. Changes in the upper bound $z_t$ are neglected because $\rho$ and $dq_v^*/dz$ are both small in the upper troposphere. The thermodynamic and dynamic contributions to $C_e$ may be written succinctly by using a mask $\mu(z)$ with $\mu = 0$ when $z<z_{\mathrm{LCL}}$ and $\mu = 1$ when $z\geq z_{\mathrm{LCL}}$:
\begin{equation}
\frac{\delta C_e}{\overline{C_e}} \approx \underbrace{\frac{1}{\overline{C_e}}\int_{0}^{z_t}\delta\left(-\left(\frac{dq_v^*}{dz}\right)_\mathrm{ma}\mu(z)\right)_e\overline{\left(\rho\tilde{w}\right)_e}\ dz}_{\text{Thermodynamic}} + \underbrace{\frac{1}{\overline{C_e}}\int_{0}^{z_t}\overline{\left(-\left(\frac{dq_v^*}{dz}\right)_\mathrm{ma}\mu(z)\right)_e}\delta(\rho\tilde{w})_e\ dz}_{\text{Dynamic}},\label{eq:thermo_dynam}
\end{equation}
where the subscript $e$ indicates that the integrand terms are averaged at and above the $99.9$th percentile of $C$. 

For extreme variables, sampling error was quantified using a block bootstrapping method. First, the samples were split into blocks of size $8$ km x $8$ km x $1$ time snapshot. Next, the set of blocks were resampled $100$ times with replacement to give $100$ new datasets of the same size. Finally, the extreme statistics (averages above a percentile) for each variable were computed for each resampling. Drawing blocks, instead of individual gridpoints, accounts for the spatially-correlated nature of heavy precipitation: convection has a larger footprint than a $1$ km x $1$ km gridbox.

\begin{figure}[ht]
	\centering\includegraphics[scale=0.8]{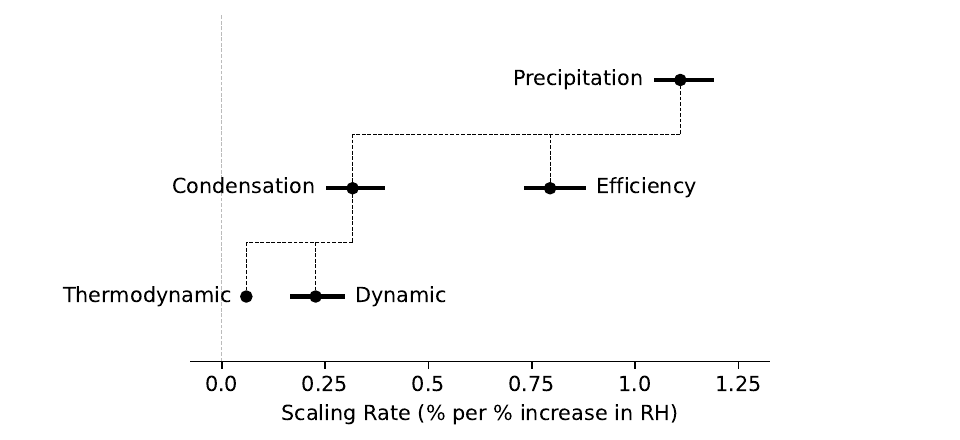}
	\caption{A physical decomposition of fractional changes in extreme precipitation (precipitation rates averaged above the 99.9th percentile) expressed as scaling rates with respect to fractional changes in RH. Variables that appear in higher rows are exactly equal to the sum of the variables in lower rows they are connected to by thin dashed lines. All scaling rates are calculated from $r_v = 0$ s m$^{-1}$ to $r_v = 2000$ s m$^{-1}$. Error bars show the 90\% confidence interval calculated using block bootstrapping of the extreme variable statistics.}\label{fig:frac_tree}
\end{figure}

Figure \ref{fig:frac_tree} shows the resulting decomposition of the $1.1\%$ per $\%$ scaling rate of $P_e$ with near-surface RH. Precipitation efficiency contributes the most to this scaling rate ($0.8\%$ per $\%$)  although fractional changes in $C_e$ ($0.3\%$ per $\%$) are also substantial. Changes in $C_e$ are further decomposed into a dynamic contribution ($0.23\%$ per $\%$) and a small thermodynamic contribution ($0.06\%$ per $\%$). Thus, all three of the thermodynamic, dynamic and precipitation efficiency changes contribute positively to the increase in $P_e$ with increasing near-surface RH.

\section{Understanding the thermodynamic contribution}\label{sec:th}

\begin{figure}[ht]
	\centering\includegraphics[scale=0.5]{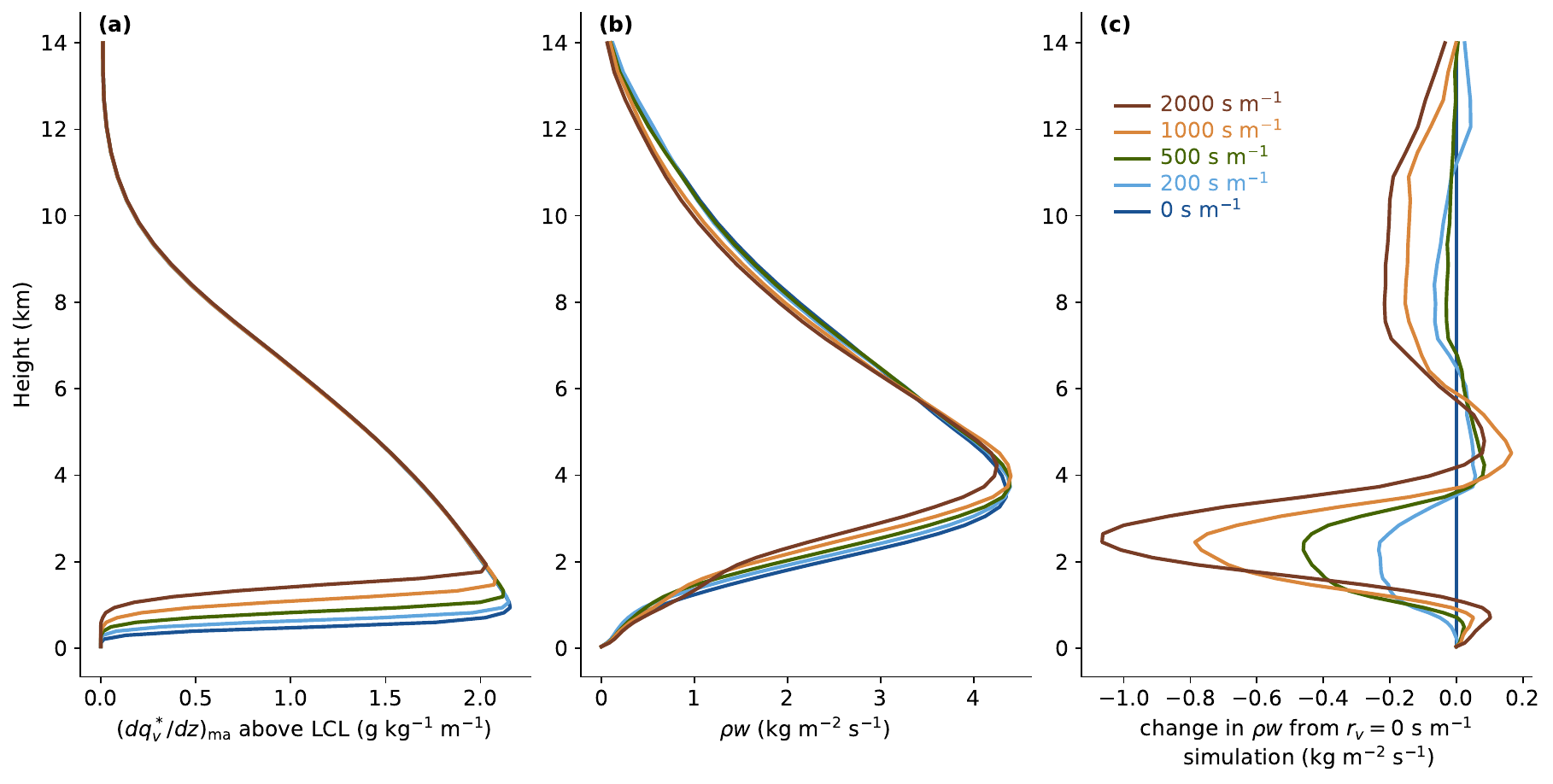}
	\caption{Vertical profiles of the factors in the integrand of the condensation integral (Equation \ref{eq:cond_int}) averaged over columns exceeding the 99.9th percentile of $C$ and plotted for simulations with different $r_v$: (a) the moist-adiabatic moisture gradient $(-dq_v^*/dz)_\mathrm{ma}$ masked to zero below the LCL, (b) the upward mass flux $\rho\tilde{w}$, and (c) the change in upward mass flux from the $r_v = 0$ s m$^{-1}$ simulation.}\label{fig:vert_cond}
\end{figure}

While small, the thermodynamic contribution of $0.06\%$ per $\%$ in Fig. \ref{fig:frac_tree} is robust with a very small error bar. Free-tropospheric temperatures are relatively horizontally homogeneous and are constrained to not change in the horizontal mean as the surface dries, and thus the thermodynamic contribution must be explained by changes in the LCL rather than changes in free-tropospheric temperatures. To illustrate this, Fig. \ref{fig:vert_cond} plots the vertical structure of the factors composing the integrand in Equation (\ref{eq:cond_int}). The moist-adiabatic moisture gradient $\left(dq_v^*/dz\right)_\mathrm{ma}$ is identical above 2 km across all values of $r_v$ (Fig. \ref{fig:vert_cond}a): temperatures above this height are held fixed by the relaxation procedure described in Section 2.2, and so $\left(dq_v^*/dz\right)_\mathrm{ma}$,  which depends on temperature and pressure, stays approximately constant with increasing $r_v$. However, since the condensation integral is not evaluated below $z_{\mathrm{LCL}}$ and the LCL rises appreciably with increasing $r_v$ (as the near-surface air dries and warms), there is some negative thermodynamic contribution below $2$ km. This contribution is small because of relatively weak values of $\overline{\rho\tilde{w}}$ this close to the surface (Fig. \ref{fig:vert_cond}b).

The thermodynamic contribution may be estimated by approximating that in the layer between LCLs, a) the cloud temperature follows a moist adiabat so that $\left(dq_v^*/dz\right)_\mathrm{ma} \approx dq_v^*/dz$, and b) the upward mass flux is approximately constant with height so that $\overline{\rho\tilde{w}}\approx M_0$ where $M_0$ is a constant. Using the definition of the LCL as the height at which lifted near-surface air becomes saturated, the thermodynamic term in Equation (\ref{eq:thermo_dynam}) is approximately
\begin{equation}
    \frac{1}{\overline{C_e}}\int_{0}^{}\delta\left(-\left(\frac{dq_v^*}{dz}\right)_{\mathrm{ma}}\mu(z)\right)\overline{\rho\tilde{w}}\ dz \approx \frac{M_0}{\overline{C_e}}\int_{z_{\text{LCL},0}}^{z_{\text{LCL},1}}\frac{dq^*}{dz}\ dz  = \frac{\overline{q_{v,s}}M_0}{\overline{C_e}}\frac{\delta q_{v,s}}{\overline{q_{v,s}}}\label{eq:thermo_est},
\end{equation}
where $q_{v,s}$ is the near-surface mixing ratio and $z_{\text{LCL},0}$ and $z_{\text{LCL},1}$ are the LCLs in the two climate states. Per Table \ref{tab:Ts}, $q_{v,s}$ decreases from $14.0$ g kg$^{-1}$ when $r_v = 0$ s m$^{-1}$ to $10.9$ g/kg when $r_v = 2000$ s m$^{-1}$: a fractional change of $0.48\%$ per $\%$ change in RH. However, with an average upward mass flux between the LCLs of $M_0 = 0.63$ kg m$^{-2}$ s$^{-1}$, $\overline{q_{v,s}}M_0 = 28.4$ mm hr$^{-1}$ is roughly a quarter of the size of the condensation rate, $\overline{C_e} = 115$ mm hr$^{-1}$.\footnote{$C$, as defined in Equation (\ref{eq:cond_int}), has units of kg m$^{-2}$ s$^{-1}$ but can be converted to mm/hr by dividing by the density of water, $\rho_w = 1000$ kg m$^{-3}$, and multiplying by 1000 to convert from m to mm.} Thus, Equation (\ref{eq:thermo_est}) predicts a thermodynamic contribution of $0.12\%$ per $\%$ increase in RH, which is double the actual thermodynamic contribution. This difference can be explained by the fact that in the layer between LCLs, $(dq^*/dz)_{\text{ma}}$ is about half as large as $dq^*/dz$ (not shown). Regardless, from Equation \ref{eq:thermo_est}, we see that the thermodynamic contribution scales at a weaker rate than the near-surface mixing ratio because the water vapor flux at cloud base ($\overline{q_{v,s}}M_0$) is much smaller than the column-integrated condensation rate.

\section{Understanding the dynamic contribution}\label{sec:dy}

The dynamic contribution to $C$ of $0.23\%$ per $\%$ increase in RH is larger than the thermodynamic contribution. With increasing $r_v$, Updrafts weaken the most in the lower troposphere, around $z=2$ to $4$ km (Figure \ref{fig:vert_cond}b). Since this is above the LCL for each simulation and $(dq_v^*/dz)_\mathrm{ma}$ is large in the lower-troposphere, the dynamic contribution is more substantial than the thermodynamic contribution. 

% \subsection{Decrease in buoyancy from upward shift of LCL}
To understand what causes this dynamic contribution, we begin by inspecting buoyancy profiles, $B(z)$, averaged over columns exceeding the $99.9$th-percentile value of $C$.  Buoyancy is defined in SAM as
\begin{equation}
	B = g\frac{T - T_{\text{env}}}{T_{\text{env}}}(1+\epsilon_v q_{v,\text{env}} - q_{n,\text{env}} - q_{p,\text{env}}) + g\epsilon_v (q_v - q_{v,{\text{env}}}) - g(q_n + q_p - q_{n,\text{env}} - q_{p,\text{env}}),\label{eq:buoyancy}
\end{equation}

where $g$ is the accelaration due to gravity, $\epsilon_v\approx 0.61$ is the ratio of water and dry air gas constants minus one, $q_n$ is the mixing ratio for non-precipitating condensates, $q_p$ is the mixing ratio for falling precipitation, and quantities with the ``env'' subscript are horizontal- and time-mean ``environmental'' profiles. SAM uses horizontal-mean environmental profiles that are time-dependent. 
%and includes a nonlinear adjustment $\epsilon_v q_{v,\text{env}}$ to the first term on the right hand side of Equation (\ref{eq:buoyancy}). 
We simplify Equation (\ref{eq:buoyancy}) by neglecting time variations of the environmental profiles. We also neglect $q_p$ and $q_{p,\text{env}}$ in our calculation of buoyancy, even though $q_p$ can contribute substantial negative buoyancy within the cloud and in the boundary layer. We justify this simplification by assuming that in columns exceeding the $99.9$th percentile of $C$, this negative buoyancy primarily acts to form and strengthen downdrafts in the lower troposphere. Just as we considered only updrafts in Equation (\ref{eq:cond_int}) and in calculating the dynamic contribution, here we consider only the buoyancy that generates those updrafts.
%can generate downdrafts, and so neglecting $q_p$ from $B$ is consistent with using $\tilde{w}$ and not $w$ in Equation (\ref{eq:cond_int}).Figure \ref{fig:buoy}a shows buoyancy and $\tilde{w}$ across our simulations. In all simulations, buoyancy peaks at around $4$ km and slowly weakens to a neutral value around $12$ km (Fig. \ref{fig:buoy}a). Buoyancy profiles are very similar to one another above each simulation's LCL. As the LCL rises, however, latent heating is lost below the LCL: as a result, drier simulations are less buoyant below their LCLs.

\begin{figure}
    \centering\includegraphics[scale=0.5]{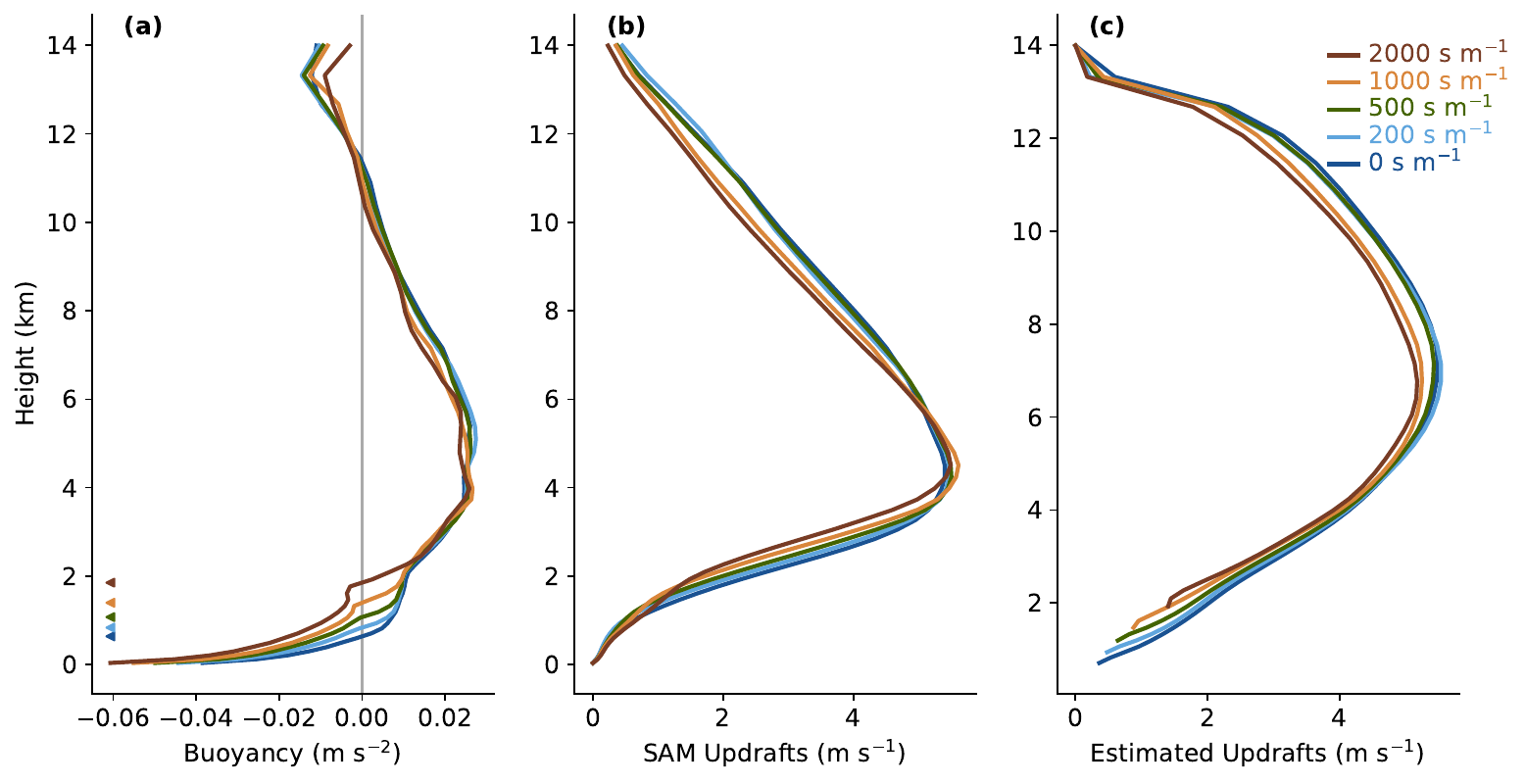}
    \caption{(a) Vertical profiles of buoyancy averaged over columns exceeding the $99.9$th-percentile of $C$. Triangles indicate the level of free convection in these columns, defined as the height where buoyancy is zero. (b) Vertical profiles of updrafts $\tilde{w}$ averaged over columns exceeding the $99.9$th percentile of $C$. (c) Vertical profiles of updrafts $w$ estimated from Equation (\ref{eq:w_JR16}). Results are shown for different values of $r_v$ as shown in the legend.}
    \label{fig:buoy}
\end{figure}

The changes in updrafts $\tilde{w}$ in the free troposphere can be broken into three distinct layers. Updrafts are weaker in drier simulations in the lower free troposphere (Fig. \ref{fig:buoy}b), consistent with the loss of positive buoyancy. In a region between roughly $4$ and $6$ km, updrafts are roughly equal across all simulations. Above $6$ km, updrafts are once again weaker in drier simulations.

The buoyancy profiles in Fig. \ref{fig:buoy}a suggest that the dynamic contribution is, like the thermodynamic contribution, mainly caused by changes in the LCL. To further investigate how buoyancy determines the updraft profiles, Fig. \ref{fig:buoy}b compares $\tilde{w}$ profiles in high-percentile $C$ columns with $w(z)$ given by solving
\begin{equation}
	\frac{1}{2}\frac{d}{dz} w^2 = aB - a(\varepsilon + b)w^2,\label{eq:w_JR16}
\end{equation}
following the recommendation of \citet{jeevanjee2016buoy}. The parameter $0\leq a \leq 1$ corresponds to the back-reaction from the environment on the parcel as it accelerates. The parameter $b > 0$ represents the effect of different types of drag on a buoyant parcel besides the entrainment of momentum. Entrainment rates $\varepsilon(z)$ are inferred from values of moist static energy in columns exceeding the $99.9$th percentile of $C$, and different entrainment profiles were used for each simulation. For the other two parameters, the same constant values of $a = 0.29$ and $b = 0.59$ km$^{-1}$ were used for all simulations. The methodology for determining the entrainment rate, $a$, and $b$ is described in the Appendix.

\begin{table}
	\begin{center}
		\begin{tabular}{c|c c c c c}
            $r_v$ (s m$^{-1}$) & $0$ & $200$ & $500$ & $1000$ & $2000$\\ \hline
            LCL (m) & $635$ & $829$ & $1059$ & $1351$ & $1724$\\
            LFC (m) & $641$ & $839$ & $1076$ & $1401$ & $1856$\\
			% $\varepsilon_0$ (km$^{-1}$) & $5.4$ & $3.3$ & $2.2$ & $1.4$ & $0.9$\\
            $w_0$ (m s$^{-1}$) & $0.3$ & $0.3$ & $0.5$ & $0.9$ & $1.4$
		\end{tabular}
		\caption{The lifted condensation level (LCL), level of free convection (LFC), and the updraft velocity $w_0$ at the LFC.}\label{tab:entr}
	\end{center}
\end{table}

Equation (\ref{eq:w_JR16}) neglects mechanical lifting (e.g., from cold pools in the boundary layer), and so to estimate $\tilde{w}$ we integrate Equation (\ref{eq:w_JR16}) upwards from the level of free convection (LFC), defined as the height where $B = 0$ in high-percentile $C$ columns (marked by triangles in Fig. \ref{fig:buoy}a). Integrating Equation (\ref{eq:w_JR16}) below the LFC introduces negative buoyancy which would decrease our estimate of $\tilde{w}$ with height in the lower troposphere. Furthermore, the LFC largely follows the LCL: the LFC is only $6$ m above the LCL in the $r_v = 0$ s m$^{-1}$ simulation and $132$ m above the LCL in the $r_v = 2000$ s m$^{-1}$ simulation (Table \ref{tab:entr}). 
Estimated $w$ profiles are initialized with an updraft velocity $w_0$ equal to $\tilde{w}$ at that simulation's LFC. Values for the LCL, LFC, and $w_0$ in each simulation are reported in Table \ref{tab:entr}.
%, by comparing updrafts from both high-percentile $C$ columns and from Equation (\ref{eq:w_JR16}) at a fixed height Fig. \ref{fig:buoy}b shows that the positive buoyancy available in wetter simulations is able to overcome this increase in $w_0$.

Updraft profiles calculated from Equation (\ref{eq:w_JR16}) capture the decreasing updraft velocities in the lower troposphere as $r_v$ increases (Fig. \ref{fig:buoy}c). The value of $w_0$ does increase with increasing $r_v$, but Fig. \ref{fig:buoy}c shows that this is overcome by the loss of buoyancy as the LCL rises, such that $w$ decreases in the lower troposphere. The updrafts
retain a memory of the loss of buoyancy from the rising LCL over a depth of no more than $(a(\varepsilon+b))^{-1}\lesssim (ab)^{-1} = 4.6$ km, and as a result differences between estimated $w$ gradually shrink and are smallest around $4-6$ km (Fig. \ref{fig:buoy}b). Above this height, a small decrease in $B$ with drier simulations causes $w$ profiles to diverge from one another again. This pattern mirrors the three-layer structure observed in $\tilde{w}$, although $w$ profiles estimated using Equation (\ref{eq:w_JR16}) are more top-heavy than $\tilde{w}$ found in high-percentile $C$ columns. This may be related to using constant values of $a$ and $b$ with height, whereas in reality these parameters may change due to e.g., a change in a buoyant parcel's aspect ratio \citep{jeevanjee2016buoy}. This may also be related to assumptions in the entrainment rate, as discussed in the Appendix.

Using $w$ estimated from Equation (\ref{eq:w_JR16}) in place of $\tilde{w}$ (except between the LCL and LFC) yields a dynamic contribution to changes in precipitation extremes of $0.23\%$ per $\%$ change in RH, which is, surprisingly, identical to the actual dynamic contribution. If we repeat the plume calculation but only allow the LFC to change (holding the buoyancy profile and $w_0$ at its $r_v=0$ s m$^{-1}$ value), the dynamic contribution is $0.33\%$ per $\%$ demonstrating that the loss of positive buoyancy from the rising LCL is more than enough to explain the dynamical contribution. Increases in $w_0$, which may be related to stronger turbulent eddies in a deeper boundary layer with a larger surface sensible heat flux, somewhat offset this loss of positive buoyancy.

\section{Understanding changes in precipitation efficiency}\label{sec:eff}

\subsection{Diagnosing contributions to precipitation efficiency}
Precipitation efficiency is nearly three times as sensitive to near-surface RH as the condensation rate (Fig. \ref{fig:frac_tree}). To understand what sets the precipitation efficiency, we take a similar approach to recent studies \citep{lutsko2018increase,da2021significant} that have estimated $\epsilon_p$ as the product of two efficiencies that respectively describe conversion of cloud condensates to precipitation ($\alpha$) and the extent to which precipitation reaches the surface without re-evaporating ($1-\beta$):
\begin{equation}
	\epsilon_p \approx \alpha(1-\beta).\label{eq:eff_apx}
\end{equation}
The ``conversion efficiency'' $\alpha \equiv A/C$ compares the  vertically integrated rate that precipitation is generated in the cloud, $A$, to the vertically integrated rate of condensation, $C$. In SAM’s one-moment microphysics scheme \citep{SAM}, two processes generate precipitation. The first process, autoconversion, activates when the mixing ratio of non-precipitating condensates $q_n$ exceeds a Kessler threshold $q_{n0}$, and takes the form\footnote{SAM defines different coefficients for different condensate phases (liquid and ice) and for different precipitation types (rain, snow, and graupel). By the phrase ``takes the form,'' we mean that SAM calculates many similar terms with $q_n$ and $q_p$ partitioned by phase and type, using the appropriate coefficients in each case, that are then summed together.}
\begin{equation}
	\left(\frac{\partial q_p}{\partial t}\right)_{\mathrm{Auto}} \propto \max(0,q_n - q_{n0}).\label{eq:auto}
\end{equation}
The second process, collection of condensates by falling precipitation, takes the form 
\begin{equation}
	\left(\frac{\partial q_p}{\partial t}\right)_{\mathrm{Accr}} \propto q_n q_p^{b_p}. \label{eq:coll}
\end{equation}
where $q_p$ is the mixing ratio for precipitating water and $b_p$ is an exponent unique to the precipitation type.
We directly saved SAM's microphysics tendencies from Equations (\ref{eq:auto}) and (\ref{eq:coll}), and used those tendencies to calculate $A$:
\begin{equation}
	A \equiv \int_0^{} \rho\left[\left(\frac{\partial q_p}{\partial t}\right)_{\mathrm{Auto}} + \left(\frac{\partial q_p}{\partial t}\right)_{\mathrm{Accr}}\right]\ dz\label{eq:A_defn},
\end{equation}

A similar approach was taken for the “sedimentation efficiency” $1-\beta$, where $\beta \equiv E/(P+E)$ measures the proportion of the generated precipitation that re-evaporates in the column at a rate $E$ and thus doesn't contribute to the surface precipitation rate $P$. In SAM's one-moment microphysics scheme, re-evaporation only occurs in locations where $q_n = 0$ and takes the form 
\begin{equation}
	\left(\frac{\partial q_p}{\partial t}\right)_{\mathrm{Evap}} \propto -f(T,q_p,\rho)(1-\mathrm{RH}(z)).\label{eq:re-evap}
\end{equation}
Note that in Equation (\ref{eq:re-evap}) $\mathrm{RH}(z)$ is evaluated at a given height (whereas elsewhere in this paper RH refers specifically to the near-surface value). 

We directly saved SAM's microphysics calculation of Equation (\ref{eq:re-evap}) and used that value to calculate $E$:
\begin{equation}
	E \equiv \int_0^{z_t} \rho\left(\frac{\partial q_p}{\partial t}\right)_{\mathrm{Evap}}\ dz\label{eq:E_defn}.
\end{equation}

The relation $\epsilon_p \approx \alpha(1-\beta)$ is exact if all precipitation generated either evaporates or reaches the surface (i.e., if $A = P+E$). However, this is only the case if we average the different terms horizontally and over a sufficiently long time (as done by \citet{lutsko2018increase}) because (a) precipitation can be transported or detrained horizontally prior to reaching the surface and (b) condensation, conversion to precipitation, evaporation, and precipitation reaching the surface can all occur at different times in a given convective lifecycle. The issue of non-locality in time is exacerbated by the use of instantaneous snapshots to calculate $P_e$ and $C_e$, since that involves no time averaging at all. 

\begin{figure}[ht]
	\centering\includegraphics[width=\linewidth]{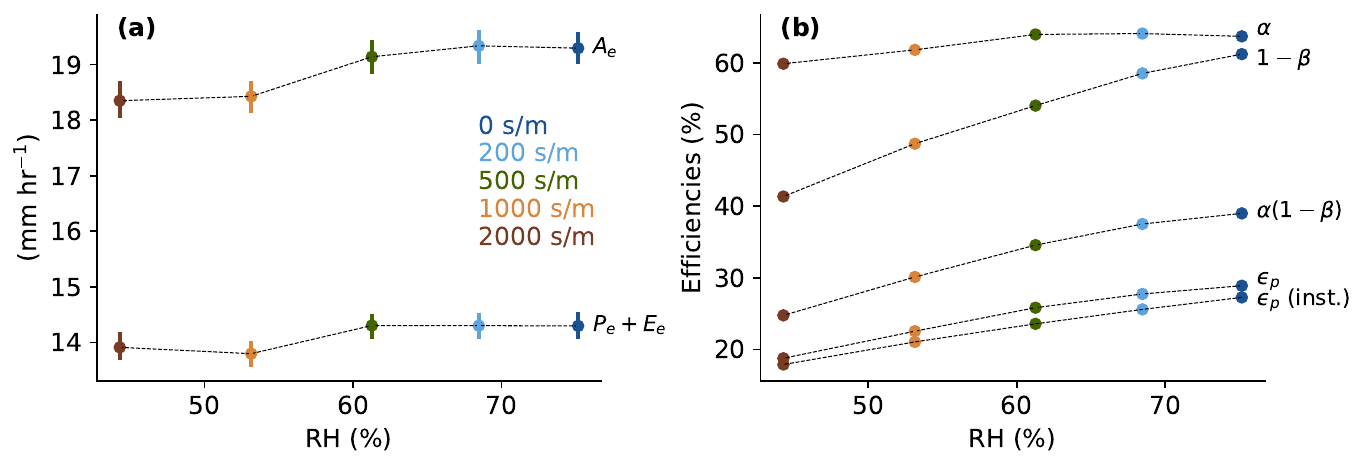}
	\caption{(a) Precipitation generation, $A_e$, and precipitation plus re-evaporation, $P_e + E_e$ as a function of RH for 3-hourly extremes averaged above the 99.9th percentile. (b) Conversion efficiency $\alpha$, sedimentation efficiency $1-\beta$, the product $\alpha(1-\beta)$, and precipitation efficiency $\epsilon_p$ as a function of $\text{RH}$, for $3$-hourly averaged precipitation extremes. Precipitation efficiency is also plotted for instantaneous (``inst.'') precipitation. In both panels, the $90\%$ confidence interval is plotted as error bars based on block boostrapping. In panel (b), these error bars span less than $2\%$.}\label{fig:residual}
\end{figure}

To mitigate the effects of non-locality in time, in this section we use $3$-hourly averaged output (instead of instantaneous snapshots) to calculate $P$, $C$, $A$, and $E$. We calculate $A_e$ as an average above the $99.9$th percentile of $A$, as done in \citet{da2021significant}. $E_e$, however, is calculated in columns that exceed the $99.9$th percentile of $P$ under the assumption that re-evaporation occurs just before precipitation reaches the surface. Figure \ref{fig:residual}a shows the residual $A_e-(P_e+E_e)$ is about $4$ to $5$ mm hr$^{-1}$ across the full range of simulations, which is roughly a quarter the size of $A_e$.

\subsection{Re-evaporation explains efficiency increase with increasing relative humidity}

Precipitation efficiency based on $3$-hourly averaged values of $P_e$ and $C_e$ closely matches the instantaneous precipitation efficiency, with both spanning about $20$ to $30\%$ across the full range of simulations (Fig. \ref{fig:residual}b). Figure \ref{fig:residual}b also shows the $3$-hourly averaged conversion efficiency $\alpha$, and the sedimentation efficiency $1-\beta$ as a function of RH. The sedimentation efficiency increases steadily with RH while the conversion efficiency is relatively constant with RH. The combination $\alpha (1-\beta)$ is close to but consistently larger than $\epsilon_p$. Much like the residual $A_e - (P_e + E_e)$, the ratio between $\epsilon_p$ and $\alpha(1-\beta)$ is an indirect measure of processes not accounted for in Equation (\ref{eq:eff_apx}), such as horizontal detrainment of falling precipitation.

\begin{figure}[ht]
	\centering\includegraphics[scale=.8]{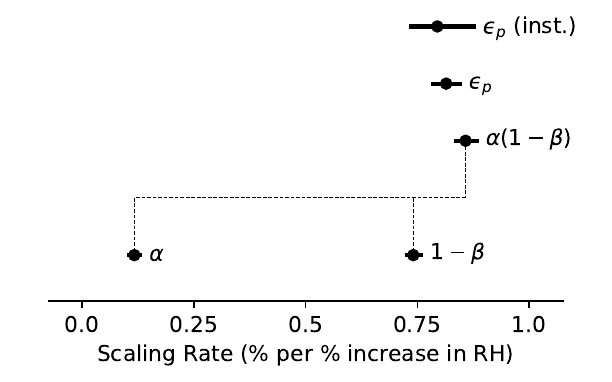}
	\caption{Decomposition of fractional changes in the $3$-hourly averaged precipitation efficiency into contributions from the conversion efficiency $\alpha$ and the sedimentation efficiency $1-\beta$. Also shown for comparison is the fractional change in instantaneous precipitation efficiency. All results are expressed as scaling rates with respect to changes in near-surface RH. The $90\%$ confidence intervals from block bootstrapping are plotted as error bars. Note that $\epsilon_p$ is not connected to the other efficiencies because $\alpha(1-\beta)$ is an approximation of $\epsilon_p$ rather than part of its decomposition.}\label{fig:eff_tree}
\end{figure}

With the use of Equation (\ref{eq:eff_apx}), fractional changes in $\epsilon_p$ may be decomposed into fractional changes in $\alpha$ and $1-\beta$, 
\begin{equation}
    \frac{\delta\epsilon_p}{\overline{\epsilon_p}} = \frac{\delta\alpha}{\overline{\alpha}} + \frac{\delta(1-\beta)}{\overline{1-\beta}},\label{eq:frac_eps}
\end{equation}
and this decomposition is plotted in Fig. \ref{fig:eff_tree}. Changes in both the conversion and sedimentation efficiencies contribute, but changes in the sedimentation efficiency are much larger. The estimate $\alpha(1-\beta)$ increases with $\text{RH}$ at a rate of $0.86\%$ per $\%$, close to $0.81\%$ per $\%$ fractional increase in $3$-hourly averaged $\epsilon_p$, implying our decomposition is reasonably accurate.

We focus on the sedimentation efficiency since its contribution is much larger. Fractional changes in sedimentation efficiency can, in turn, be attributed to fractional changes in  $A$ and $E$:
\begin{equation}
	% \frac{\delta\alpha}{\overline{\alpha}} &= \frac{\delta A}{\overline{A}} - \frac{\delta C}{\overline{C}},\label{eq:frac_alpha}\\ 
	\frac{\delta(1-\beta)}{\overline{1-\beta}} \approx \frac{\overline{E}}{\overline{P}}\left(\frac{\delta A}{\overline{A}} - \frac{\delta E}{\overline{E}}\right),\label{eq:frac_beta}
\end{equation}
where we have substituted in $\overline{1-\beta} = \overline{P}/\overline{A}$ and $\delta(1-\beta) = \delta(1 - E/A)$. Equation (\ref{eq:frac_beta}) is approximate because both of these substitutions  assume that $P=A-E$ exactly. Equation (\ref{eq:frac_beta}) states that the fractional change in $1-\beta$ is set by a balance between fractional changes in the amount of precipitation generated, $A$, and the amount of precipitation that re-evaporates while falling, $E$. 

\begin{figure}[ht]
	\centering\includegraphics[scale=0.7]{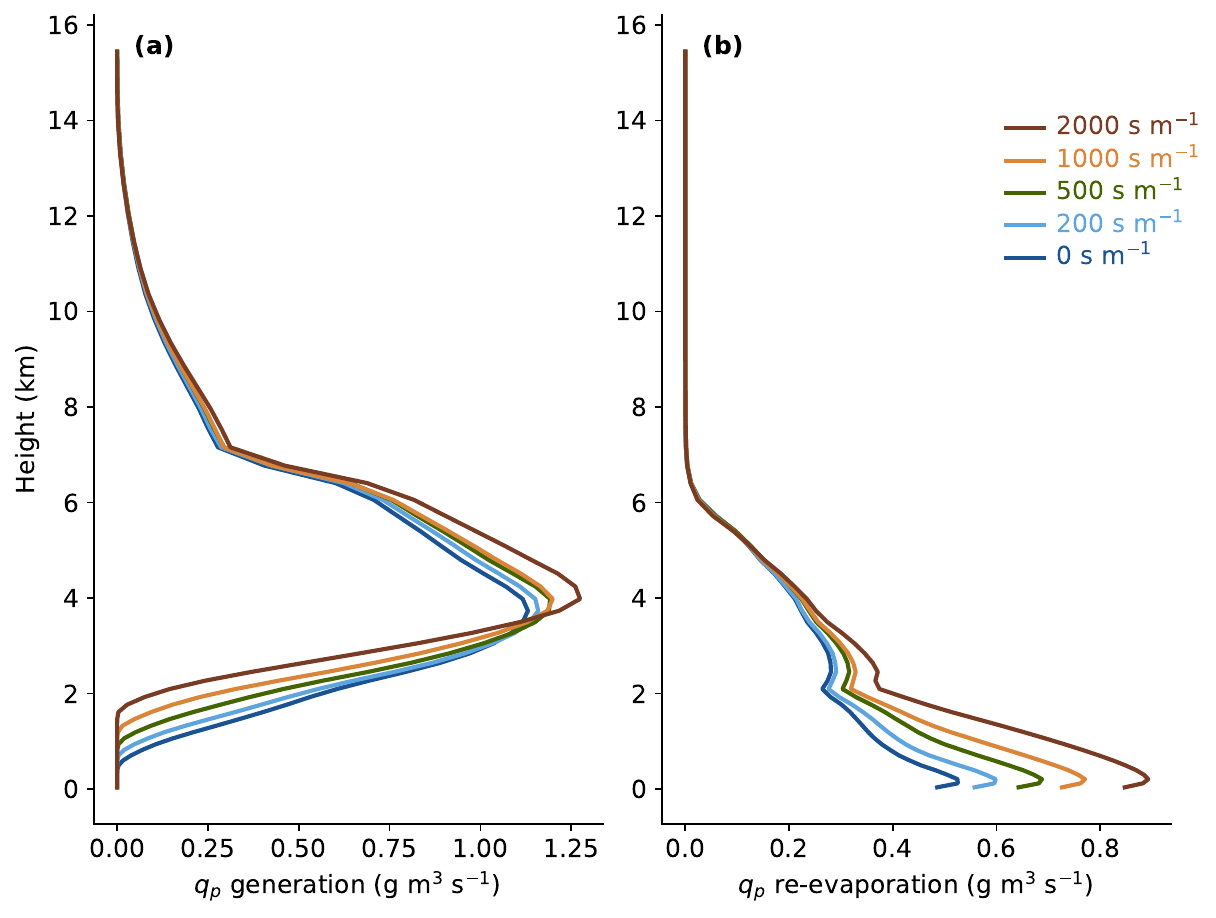}
	\caption{Vertical profiles of (a) the precipitation re-evaporation rate (the integrand of Equation (\ref{eq:E_defn})) and (b) the precipitation generation rate (the integrand of Equation (\ref{eq:A_defn})) for extremes averaged above the 99.9th percentile in simulations with varying $r_v$ (see legend).}\label{fig:vert_eff}
\end{figure}

The contributions to $A_e$ and $E_e$ at different vertical levels across simulations with various $r_v$ are shown in Fig. \ref{fig:vert_eff}. As $r_v$ increases, the contribution to $A_e$ increases in the mid-troposphere but decreases in the lower free troposphere, while the contribution to $E_e$ increases throughout the boundary layer and lower free troposphere. Increases in $E_e$ with $r_v$ are a direct result of a deeper and drier boundary layer at high $r_v$, since re-evaporation is proportional to $1-\mathrm{RH}(z)$ per Equation (\ref{eq:re-evap}) and is only non-zero outside of the cloud. When integrated vertically, these profiles yield an $A_e$ that decreases relatively slowly with increasing $r_v$ and an $E_e$ that increases with increasing $r_v$. Thus changes in both $A_e$ and $E_e$ contribute to a decrease in $1-\beta$ as $r_v$ is increased. 

\begin{figure}[ht]
    \centering
    \includegraphics[scale=0.6]{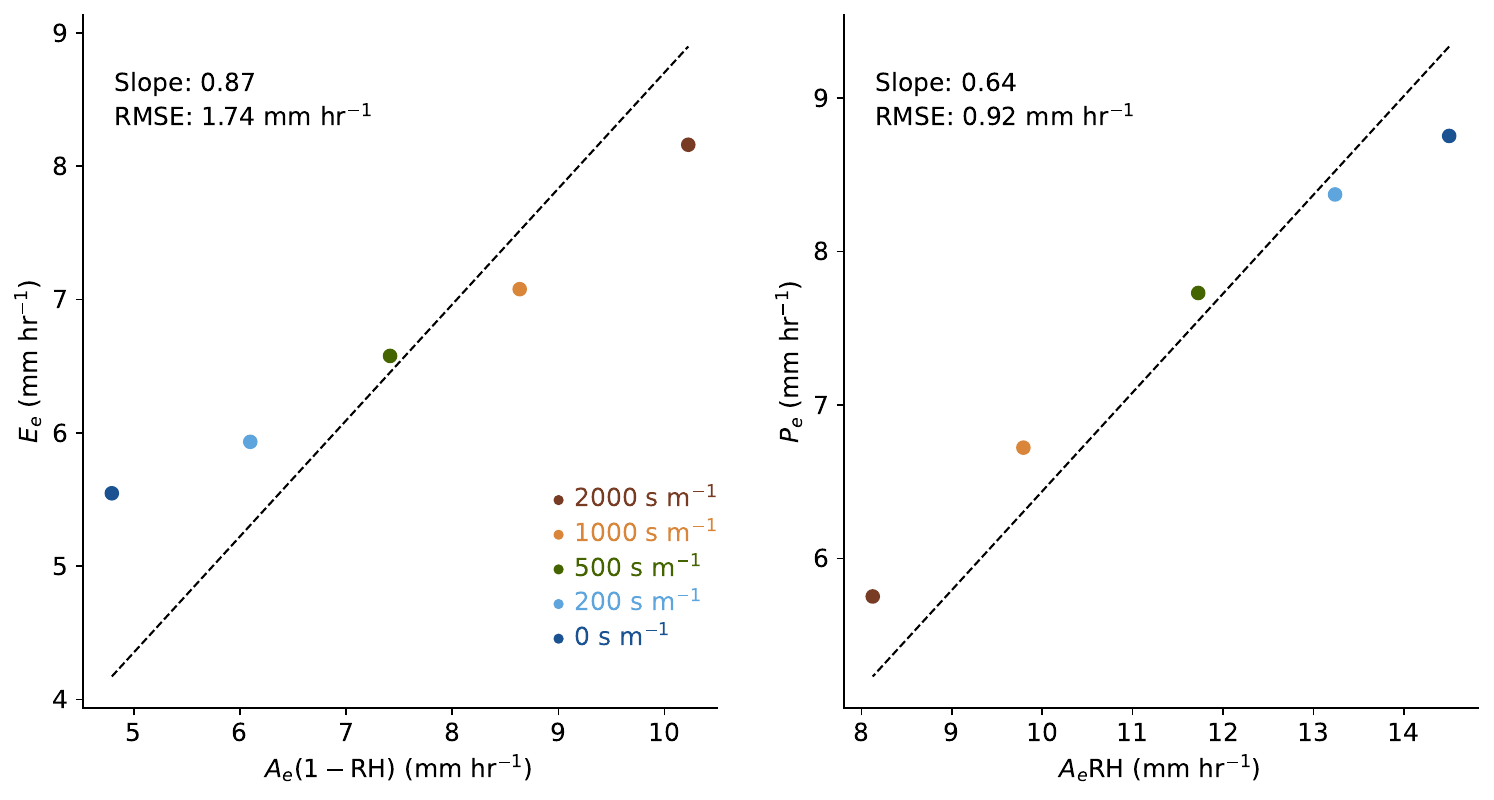}
    \caption{(a)  $E_e$ versus $A_e(1-\mathrm{RH})$ and (b) $P_e$ versus $A_e\mathrm{RH}$ for simulations with varying $r_v$ (legend). Dashed lines show linear least-squares regression fits with an intercept forced to equal zero, and in each panel the slope and root-mean square error (RMSE) of the fit is reported.}
    \label{fig:empirical}
\end{figure}

Finally, we derive some scalings that show how sedimentation efficiency might be related to changes in near-surface RH. Given that evaporation in the subcloud layer will roughly scale with the amount of precipitation generated at higher vertical levels, and given the dependence of Equation (\ref{eq:re-evap}) on $1-\mathrm{RH}(z)$, we approximate $E\sim A (1 - \mathrm{RH})$ such that $\beta = E/A \sim (1-\mathrm{RH})$. Substituting this into Equation (\ref{eq:frac_beta}) gives that
\begin{equation}
    \frac{\delta(1-\beta)}{\overline{1-\beta}} = -\frac{\overline{E}}{\overline{P}}\frac{\delta(1-\mathrm{RH})}{\overline{1-\mathrm{RH}}},\label{eq:E_scale_apx}
\end{equation}
which directly relates changes in the sedimentation efficiency to changes in RH. The approximation $E_e \sim A_e(1-\mathrm{RH})$ underestimates changes in $E_e$ (Fig. \ref{fig:empirical}a), likely because it neglects the vertical variations of $RH(z)$ and the precipitation generation rate, as well as the detailed dependence of evaporation rate on terminal velocity and other microphysical factors. 

An alternate but related scaling is $P\sim A \, \mathrm{RH}$.\footnote{In the special case that these two scalings have slopes of $1$, i.e. that $E \approx A(1-\mathrm{RH})$ and $P \approx A\,\mathrm{RH}$, these two scalings are identical so long as $P+E\approx A$.} This scaling is simple and intuitive: $P\sim A \, \mathrm{RH}$ states that precipitation at the surface a) is directly proportional to the amount of condensation converted into precipitation aloft, and b) that drier boundary layers reduce surface precipitation (implicitly by increasing re-evaporation). Figure \ref{fig:empirical} shows that $P_e \sim A_e\mathrm{RH}$ is a better empirical fit than $A_e(1-\mathrm{RH})$ is to $E_e$. Using $A_e \approx P_e + E_e$ we have the simple prediction that $1-\beta \sim \mathrm{RH}$ and
\begin{equation}
    \frac{\delta(1-\beta)}{\overline{1-\beta}} = \frac{\delta\mathrm{RH}}{\overline{RH}},\label{eq:P_scale_apx}
\end{equation}
i.e., that $1-\beta$ scales with RH at a rate of $1\%$ per $\%$. Equation (\ref{eq:P_scale_apx}) is a modest overestimate of the actual scaling rate of sedimentation efficiency at $0.8\%$ per $\%$. 

Overall, these simple scalings are approximate but help by showing how the sedimentation efficiency and precipitation rate can be related to near-surface RH.

\section{Conclusions}
\label{sec:conclusions}

Using a CRM run to states of RCE, we found that convective precipitation extremes are sensitive to near-surface RH: between our wettest and driest simulations, instantaneous precipitation extremes fractionally decrease by $1.1\%$ for every $1\%$ fractional decrease in near-surface RH. When normalized by absolute rather than fractional changes in RH, precipitation extremes are more sensitive to near-surface RH over drier surfaces. Specifically, scaling rates range from $1.4\%$ per $\%$pt between our two wettest simulations ($\overline{\mathrm{RH}} = 72\%$) to $2.5\%$ per $\%$pt between our two driest simulations ($\overline{\mathrm{RH}} = 49\%$). 

\begin{figure}[ht]
	\centering\includegraphics[scale=0.55]{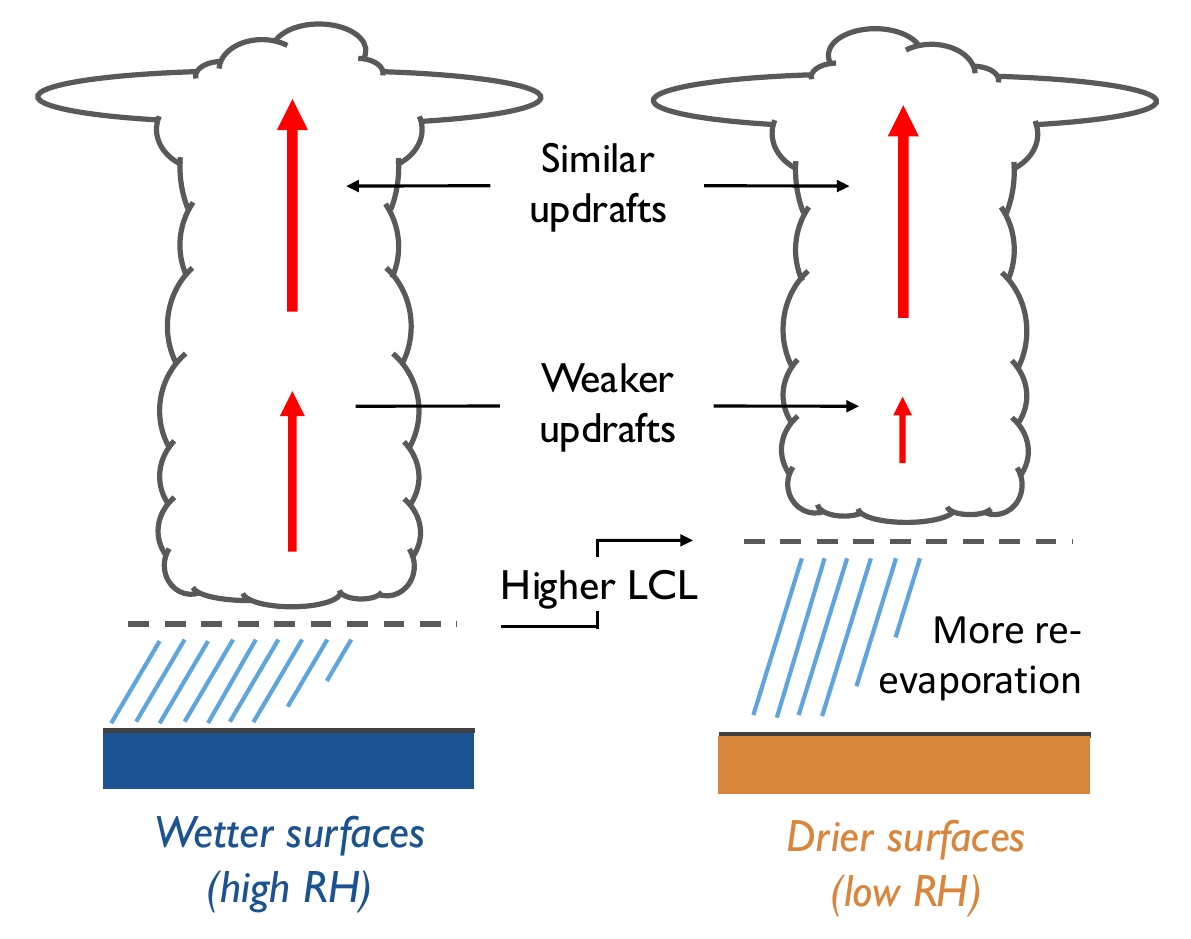}
	\caption{A schematic showing three responses of the intensity of precipitation extremes to lower RH found over a less evaporative surface. First, there is a direct thermodynamic response to the higher LCL assuming condensation only occurs above the LCL. Second, updrafts (red arrows) are weaker in the lower troposphere, also associated with the higher LCL, because rising parcels gain positive buoyancy above cloud base. Third, re-evaporation of precipitation is greater due to a decrease in RH in the deeper sub-cloud layer. Each of these changes weakens precipitation intensity over the drier surface.}\label{fig:schematic}
\end{figure}

Three distinct physical mechanisms, all associated with changes in near-surface RH, explain these scaling rates. First, a weak thermodynamic contribution is found in direct response to changes in the LCL, which follow from changes in near-surface RH (Section \ref{sec:th}). Second, a dynamic contribution also depends on changes in the LCL because positive buoyancy is only realized above the cloud base (Section \ref{sec:dy}). Third, re-evaporation is proportional to a factor of $1-\mathrm{RH}(z)$, and so precipitation efficiency is much lower in simulations with deeper, drier boundary layers (Section \ref{sec:eff}). These effects are illustrated schematically in Fig. \ref{fig:schematic}.

The above three physical mechanisms-- involving changes in the LCL and changes in re-evaporation-- are all distinct from mechanisms that have been used to explain the Clausius-Clapeyron scaling of precipitation extremes with warming. Clausius-Clapeyron scaling has been explained by relating precipitation extremes to near-surface temperatures via specific humidity, either through moisture convergence arguments \citep{trenberth1999framework,allen2002constraints} or through simplifications of the condensation integral \citep{ogormanschneider,abbott2020convective}. All of these explanations rely on an assumption of constant RH, but the scaling rates calculated in this study suggest that a decrease in RH of $2.5\%$ over a dry surface, or a decrease in RH of $4\%$ over a moist surface, is sufficient to offset the effect of 1 K of warming and a Clausius-Clapeyron scaling rate of $\sim 6\%$ K$^{-1}$. Note our simulations already include the effect of surface warming as the surface dries when the free-tropospheric temperature is held constant (Fig. \ref{fig:th_rh}b). This warming, explained by the top-down model of the land-ocean warming contrast \citep{joshi2008contrast,byrneogormancontrast}, would be in addition to the 1 K of surface warming mentioned above that warms the free troposphere (and thus increases $(dq_v^*/dz)_{\text{ma}}$ in Equation (\ref{eq:cond_int})).

Overall, despite many differences including considering instantaneous or 3-hourly precipitation rather than daily precipitation, our study provides support, based on convection-resolving simulations, for the conclusion of \citet{williams2022summer} that decreases in relative humidity are important for changes in summertime midlatitude precipitation extremes over land. \citet{williams2022summer} considered whether such a correlation may be related to the dependence of convective inhibition (CIN)-- negative buoyancy in the lower troposphere-- on RH, since \citet{chen2020cin} found that CIN increased as RH decreased and the LCL and LFC rose. While \citet{williams2022summer} did find a correlation between changes in seasonal-mean CIN and the dynamical contribution to changes in precipitation extremes, the correlation was much weaker for CIN on the day of the precipitation event, casting doubt on the causality. Our results suggest an alternative cause: that updrafts weaken because of a decrease in positive buoyancy rather than an increase in negative buoyancy. However, CIN plays less of a role for convection in RCE compared to convection over midlatitude land \citep[e.g.,][]{markowskirichardson,agardemanuel,emanuelcape}, and thus further investigation of the midlatitude land case is warranted.

We find that the dynamic contribution is smaller than the contribution from changes in precipitation efficiency,  whereas \citet{williams2022summer} found a substantial dynamical contribution and did not consider a contribution from changes in precipitation efficiency.  
However, re-evaporation within downdrafts may be indirectly represented in the dynamical contribution of \citet{williams2022summer} because they evaluate the condensation integral with a lower bound of $z=0$ instead of $z = z_{\mathrm{LCL}}$ and allow for $w < 0$. Their definition is the same as the condensation integral introduced by \citet{ogormanschneider}, but in this paper we use a different definition (Equation (\ref{eq:cond_int})) that measures only condensation driven by updrafts above the cloud base. As a result of our alternate definition, we were able to separately diagnose changes in the precipitation efficiency resulting from changes in re-evaporation (Section \ref{sec:eff}).

Our finding that decreases in RH weaken precipitation extremes seems to be at odds with a scaling analysis of observed variability by \citet{lenderink2024} which found that precipitation extremes were stronger at lower RH for a given dewpoint temperature. Part of this result of \citet{lenderink2024} is a statistical effect related to conditioning on wet hours only, but the result still persisted more weakly when all hours were considered. One possible reason for the discrepancy is that our simulations focus on changes in horizontal- and time-mean near-surface RH in a state of RCE whereas \citet{lenderink2024} analyze weather variability in hourly near-surface RH. Such weather variability could allow environments with lower near-surface RH to correspond to greater convective instability or greater convective organization.

\citet{skinner17} found in GCM simulations that stomatal closure causes widespread decreases in RH over land and decreases in mean precipitation in northern midlatitudes in summer, but that stomatal closure could actually increase mean and extreme precipitation in some regions of the deep tropics over land. These contrasting precipitation sensitivities in different regions suggest that large-scale dynamics may play an important role in the response of precipitation extremes to near-surface $\text{RH}$. Our results could be extended to include the influence of changing large-scale vertical velocity in future work by using a parameterization of the large-scale dynamics such as the weak-temperature gradient approximation \citep{sobel2000modeling,raymond2005modeling}.

In conclusion, we have shown how changes in near-surface RH affect the intensity of precipitation in the simplest statistical-equilibrium case of RCE. In future work, we plan to address the scaling of precipitation extremes across a wide range of different temperatures and humidities in a similar RCE setting. In addition to incorporating the role of large-scale circulation as discussed earlier, future work should also include simulations with a diurnal cycle, since convection over land is heavily influenced by the diurnal cycle. It would also be interesting to consider the case of organized convection which may respond differently to changes in surface RH. 

\appendix
\appendixtitle{Determining parameters in the plume vertical velocity equation}

\begin{figure}[ht]
    \centering
    \includegraphics[scale=0.5]{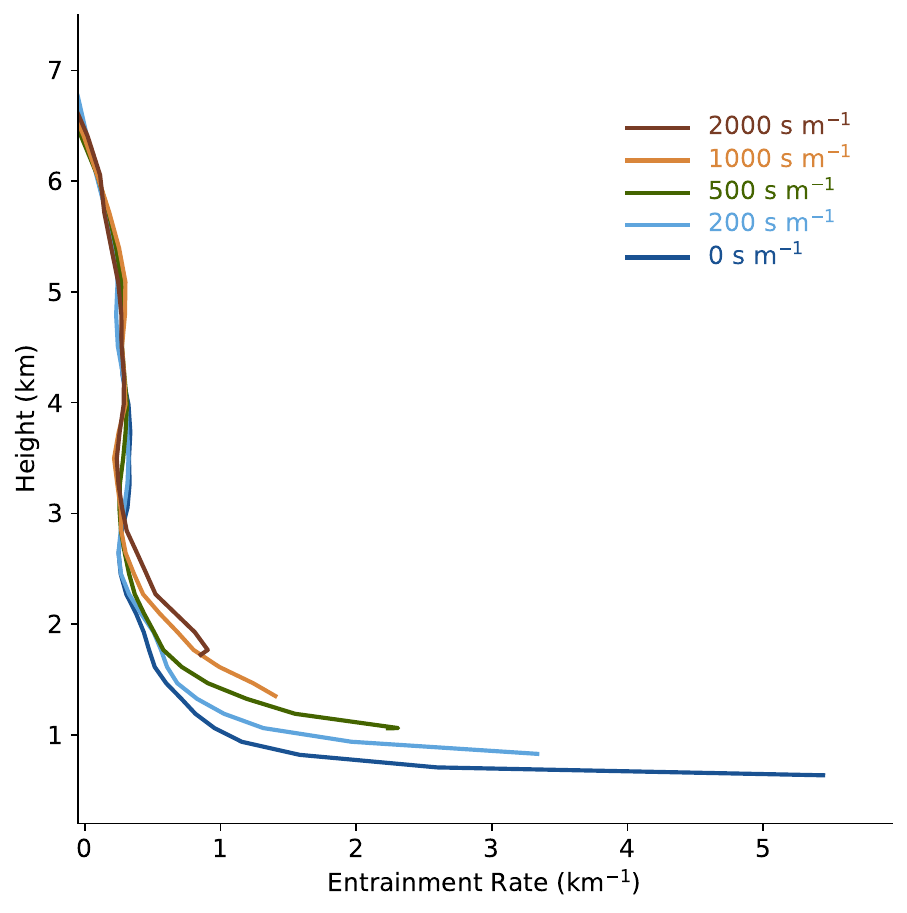}
    \caption{Entrainment rates from extreme $C$ columns (averaged above the 99.9th percentile) calculated by solving for $\varepsilon$ in Equation (\ref{eq:mse_entr}) for simulations with different $r_v$ (see legend).}
    \label{fig:entr_decay}
\end{figure}

The entrainment rates used in Section (\ref{sec:dy}) were estimated by assuming that within extreme $C$ columns (averaged above the 99.9th percentile), saturation frozen moist static energy (MSE) $h^*$ is mixed with the environmental frozen MSE $h_\text{env}$ following a bulk entraining plume:
\begin{equation}
    \frac{dh^*}{dz} = -\varepsilon(z)(h^* - h_\text{env}).\label{eq:mse_entr}
\end{equation}
Frozen MSE is defined following SAM thermodynamics as
\begin{equation}
    h = c_p T + gz + L_v q_v - L_f q_n (1-\omega(T)),\label{eq:frozen_MSE}
\end{equation}
where $c_p$ is the specific heat at constant pressure, $g$ is gravity, $L_v$ and $L_f$ are the latent heats of vaporization and fusion, respectively, and $\omega(T)$ is a partition function that SAM uses to distinguish between liquid and ice condensates \citep{SAM}. $h^*$ is defined by replacing $q_v$ with $q_v^*$ in Equation (\ref{eq:frozen_MSE}). Based on this assumption, $\varepsilon$ can be computed by inverting Equation (\ref{eq:mse_entr}):
\begin{equation}
    \varepsilon = -\left(\frac{1}{h^* - h_{\text{env}}}\frac{dh^*}{dz}\right).
\end{equation}
Figure \ref{fig:entr_decay} plots the entrainment rates estimated from high-percentile $C$. Above $6$ km, $h^*$ increases with height in high-percentile $C$ columns, which implies an unphysical $\varepsilon < 0$. Increasing $h^*$ with height cannot be explained by a single entraining plume, but it can be explained by a spectrum of plumes with different entrainment rates \citep{zhou2019spectrum}. It is plausible that a spectral approach may improve upon the $w$ estimated from Equation (\ref{eq:w_JR16}), which is too strong in the region where $h^*$ increases with height. However, given our interest specifically in the lower free troposphere, we use a single bulk plume for its simplicity and set $\varepsilon = 0$ km$^{-1}$ where it would otherwise be negative. This is a reasonable simplification to the extent that $\varepsilon \ll b$ in the upper troposphere.

Constant values for the parameters $a$ and $b$ were fit to the $r_v = 0$ s m$^{-1}$ simulation and were determined in two steps.
First, $b$ was calculated using values for $\tilde{w}$, $B$, and $\varepsilon$ at the height $z_{\mathrm{max}}$ where $\tilde{w}$ achieves its maximum value, $\tilde{w}_{\mathrm{max}}$. At this height, the left hand side of Equation (\ref{eq:w_JR16}) vanishes, so that
\begin{equation}
    b = \frac{B(z_{\mathrm{max}})}{\tilde{w}^2(z_{\mathrm{max}})} - \varepsilon(z_{\mathrm{max}}).
\end{equation}
To determine $a$, we solved Equation (\ref{eq:w_JR16}) for a range of $a$ values between $0$ and $1$, using this value of $b$. We chose the value of $a$ that matched the maximum value of $w$ to the high-percentile $C$ column's maximum $\tilde{w}$. When choosing $a$, we did not require that $w$ achieved its maximum value at the same height $z_{\mathrm{max}}$ as in the high-percentile $C$ column.

\acknowledgments
We thank Tim Cronin for helpful discussions, particularly regarding the analysis of precipitation efficiency. This research is part of the MIT Climate Grand Challenge on Weather and Climate Extremes. Support was provided by Schmidt Sciences, LLC.

\datastatement
SAM is available at: http://rossby.msrc.sunysb.edu/SAM.html.
Our code and data, including modifications to SAM, scripts to generate our simulations, post-processed data, and scripts and Jupyter notebooks to analyze data are available at: https://doi.org/10.5281/zenodo.14416632

\bibliographystyle{ametsocV6}
\bibliography{refs}

\end{document}